\documentclass[ 12pt, reqno]{amsart}
\usepackage{amssymb,latexsym}

\begin{document}
\begin{center} Derivation of Unified Analytic Electron Interaction Integrals over Slater Orbitals for Diatomic Molecules \end{center}
 \begin{center} E.V.Rothstein \end{center}
\begin{center} evrothstein@gmail.com  \end{center}  
  
In previous work, the necessary integrals arising from correlated wavefunctions were expressed in forms suitable for numerical integration.  For the evaluation of $r_{12}$ and 1/$r_{12}$  integrals an analytical formula is derived.  Combinatorial aspects of more-than-two-electron integrals are discussed.

\begin{center} INTRODUCTION \end{center}

For the two-center problem we have thus far $ \text{published}^{1} $ results showing that the correlation integrals can be expressed as iterated one-dimensional integrals over certain functions. If one expresses $r_{12}$ as  $r_{12}^{2}$  multiplied by 1/$r_{12}$ ; all the integrals are expressible in terms of two functions.  One is $K^{m}_{l,a} (z) $, the definite integral over the one-electron elliptical orbital labelled a. The other is 
\begin{equation*}F^{m_{1}(t_1),m_{2}(t_{2})}_{l_{1},l_{2}} (z)  = \frac{ -d}{dz} \left[   \left( \frac{ Q^{m_1}_{l_{1}}(z) }{ P ^{m_1}_{l_{1}}(z)    }   \right)^{t_1}  \left(  \frac{ Q^{m_2}_{l_{2}}(z) }{ P ^{m_2}_{l_{2}}(z)    }  \right)^{t_2}          \right] \end{equation*} 
where the exponents $t_i$ are zero or one, denoting the presence or absence of that term.  The $P^{m}_{l} (z)$, $Q^{m}_{l} (z)$ are associated Legendre polynomials of the first and second kinds.  To numerically integrate the matrix element  $\left< \frac{1}{r_{ij}  r_{kl} r_{mn}} \right>$ one calculates the K and F functions at the appropriate mesh points, stores them in the computer and uses them over and over again in making the integrals.  The proceedural schema for combining the $K^{m}_{l,a}(z)$ and $F^{m_{1}(t_{1}),...etc.}_{l_{1},...etc.} (z)$  values becomes more complicated as the number of $r_{ij}$ terms in the integrand increases.  The results of graph theory (which we will discuss later ) could be of some use in knowing which integrals to evaluate.
    Two-electraon hybrid and coulomb $\text{integrals}^{2-8}$ are expressible as a finite sum of exponential type integral functions and spherical Bessel type functions.  The exchange integrals can also be evaluated by using the $\text{Neumann}^{9}$ expansion in elliptical coordinates and integration by parts technique of $\text{Podolanski}^{10}$ , with a less complicated result than that arrived at by Silverstone and $\text{Kay}^{11}$ ( using a more generall method).  The expression is an infinite sum in terms of the exponential type integral function and spherical Bessel type functions. The question of the convergence of the ininite sum has been studied by Kolos and $\text{Wolniewicz}^{12}$.
   Zivkovic and $\text{Murrell}^{13}$ have presented a method for the special case ( axially symmetric orbitals with equal screening constants ) which, according to their claim, appears to be faster than previously available methods.  The present work has generalized that method.

\begin{center} TWO-CENTER GEOMETRY and FORMULA \end{center}
     
The z axis goes in the direction from center A to center B ;  $\phi$ is the angle around the z  axis by the right hand rule.  $r_{ai}$ is the distance from A to particle i ; $\theta_{ai}$ is the angle between the z axis and $r_{ai}$ with direction so that   z crossed with  $r_{ai}$ is out of the page.  Similar definitions for $r_{bi}$ and  $\theta_{bi}$.  R is the distance between center A and center B.  
\begin{equation*}  r_{ai} = \frac{R}{2} ( \xi_{i} + \eta_{i} ) ; \quad  \quad cos(\theta_{ai}) = \frac{\xi_{i} \eta_{i} + 1 }{\xi_{i} + \eta_{i} }       \end{equation*}

\begin{equation*}  r_{bi} = \frac{R}{2} ( \xi_{i} - \eta_{i} ) ; \quad  \quad cos(\theta_{bi}) = \frac{\xi_{i} \eta_{i} - 1 }{\xi_{i} - \eta_{i} }       \end{equation*}

\begin{equation*}  ( \quad  \xi_{i} \geq 1 ;  \quad \quad -1 \le \eta_{i} \le 1 \quad ) \end{equation*}

\begin{equation*} P_{l}^{m} ( \eta) = \frac{ ( - )^{m} ( 1 - \eta^2)^{m/2}} {2^{l} l! } \frac{ d^{ l + m}}{d\eta^{ l+m}} (\eta^2 - 1 )^{l}     \end{equation*}
( Ref. 14, 8.6.6 and 8.6.18 )

\begin{equation*} \frac{d^{ l+m} }{d\eta^{ l+m}} ( \eta^{2} - 1)^{l} = \sum_{s=0}^{ l } \binom{l}{s}  \frac{ ( - )^{s} ( 2 l - 2 s ) !  }{ ( 2 l - 2 s - l - m ) ! } \eta^{ 2 l - 2 s - l - m }      \end{equation*}
  provided   $ l - m \geq   2 s $

\begin{multline*} P_{l}^{m} ( cos \theta_{c}) = \frac{ ( - )^{m}( \xi^{2} -1 )^{m/2}  ( 1 - \eta^2)^{m/2}} {2^{l} (\xi \pm \eta )^{ l }  }  \\   \times \sum^{[\frac{ l - m }{2}]}_{s=0}  \frac {( - )^{s} ( 2 l - 2 s )! ( \xi \eta \pm 1 )^{ l - m - 2 s } ( \xi \pm \eta )^{ 2 s } } { s! (l - s )! ( l - 2 s -m )!}      \end{multline*}
Here   c  denotes  center a  with upper sign  or center b  with lower sign .     ( Ref. 7, II20, II21 ) 

\begin{multline*}  ( \xi \pm \eta )^{ 2 s } ( \xi \eta \pm 1 )^{ l - m - 2 s } = \sum_{ p = 0 }^{ 2 s } \sum_{ q = 0 }^{ l - m - 2 s } \binom{ l - m - 2 s }{ q } \binom{ 2 s }{ p }   \\  \times  ( \pm )^{ l - m - 2 s - q + p } \xi^{2 s - p + q } \eta^{ p + q }     \end{multline*}
   
 \begin{multline*} \Phi_{  c }  = ( - )^{ ( m - | m | ) / 2 }  ( 2 \delta )^{ n + \frac{ 1 }{ 2 } }   \left[ \frac{ ( 2 l + 1 ) ( l - | m | ) ! }{ 4 \pi ( 2 n ) ! ( l + | m | ) ! } \right]^{ \frac{ 1 }{ 2 } }    \\    \times    e^{ i m  \phi } r_{  c }^{ n - 1 }  e^{ - \delta  r_{  c }}  P^{ | m | }_{ l } ( cos \theta_{  c } )    
   \end{multline*}   

$\Phi_{c} $ is a normalized Slater orbital in spherical coordinated  centered on center a or center b with quantum numbers ( n, l, m ) and screening constant $\delta $ (not the usual Kronecker delta function).  This can be transformed into the two-center coordinate system for orbital centered on a with the upper sign (positive) and for b with the lower sign (negative).

\begin{multline*}  \Phi_{  c }  =   ( 2 \delta )^{ n + \frac{ 1 }{ 2 } }   \left[ \frac{ ( 2 l + 1 ) ( l - | m | ) ! ( l + | m | )!}{ 4 \pi ( 2 n ) !  } \right]^{ \frac{ 1 }{ 2 } }    \\    \times    e^{ i m  \phi } \left( \frac{ R }{ 2 } \right)^{ n - 1 } ( \xi \pm \eta )^{ n - l - 1 } e^{ - \delta R ( \xi \pm \eta) /2 } (\xi^2 - 1 )^{ |m|/2 }( 1 - \eta^2)^{ | m | /2 }  \\ \times \sum_{ s = 0 }^{[ \frac{ l - | m | }{ 2 }] }  \frac{ ( - )^{ s + ( m + | m | ) / 2 } }{  2^{ l } l! }  \binom{ 2 l - 2 s }{ l - | m | - 2 s } \binom{ l }{ s }  \\ \times  \sum_{ p = 0 }^{ 2 s } \sum_{ q = 0 }^{ l - | m | - 2 s } \binom { l - | m | - 2 s }{ q } \binom{ 2 s }{ p } \\ \times  ( \pm )^{ l - | m | - q + p } \xi^{ 2 s - p + q } \eta^{ p + q } 
 \end{multline*}

\begin{multline*}    \frac{ 1 }{ r_{ 1 2 } } = \frac { 4 }{ R } \sum_{\mu = 0 }^{ \infty } \sum_{ \sigma = - \mu }^{ \mu } ( - ) ^{ \sigma }  \left( \frac{ 2 \mu + 1 }{ 2 } \right) \left[ \frac{ ( \mu - | \sigma | ) ! }{ ( \mu + | \sigma | ) !}   \right]^{ 2 } \\ \times P_{\mu}^{|\sigma|}(\xi_{1 < 2}) Q_{\mu}^{|\sigma|}(\xi_{ 2 > 1 } ) P_{ \mu }^{|\sigma |} (\eta_{1}) P_{\mu}^{|\sigma|}(\eta_{2}) e^{ i \sigma ( \phi_{1} - \phi_{2} ) }    \end{multline*}
 
In the Neumann expansion ( ref. 9 ) for the inverse of the  distance between particles 1 and 2, the notation 
 $ P_{\mu}^{|\sigma|}(\xi_{1 < 2}) Q_{\mu}^{|\sigma|}(\xi_{ 2 > 1 } ) $
means $ P_{\mu}^{|\sigma|}(\xi_{1 }) Q_{\mu}^{|\sigma|}(\xi_{ 2  } )$ if $\xi_1 < \xi_2$ and 
$ P_{\mu}^{|\sigma|}(\xi_{ 2}) Q_{\mu}^{|\sigma|}(\xi_{  1 } )$  if $\xi_1 > \xi_2$

$d \tau $ = volume element = $d \tau_{1} d \tau_{2} $
\begin{equation*} d\tau_{i} = \left( \frac{R}{2} \right)^{3}  ( \xi_{i}^{2} -   \eta_{i}^{2} ) d \xi_{i} d \eta_{i} d \phi_{i}   \end{equation*}

 \begin{equation*} \text{Exchange integral } = I = \int \int d \tau_{1} d \tau_{2}  \Phi_{ 1 a } ( 1 ) \Phi_{ 3 b } ( 1 ) \frac{ 1 }{ r_{ 1 2 } } \Phi_{ 2 a } ( 2 ) \Phi_{ 4 b } ( 2 )   \end{equation*}

\begin{multline*} \text{Integrating over the azimuthal angles : }\\  \sum_{ \sigma = - \mu }^{ \mu } \int_{ 0 }^{ 2 \pi } \int_{0}^{ 2 \pi } d \phi_{1} d \phi_{2}  e^{ i ( m_{1} + m_{3} ) \phi_{ 1 } } e^{i ( m_{ 2 } + m_{ 4 } ) \phi_{ 2 } }  e^{i \sigma ( \phi_1 - \phi_2 ) }      \\              =  ( 2 \pi )^{2} \delta( \sum m_{i} ) \delta ( |\sigma| - | m_{2} + m_{4} | )  \end{multline*}

\begin{multline*}  ( 1 - \eta_{ 1 }^{ 2 }  )^{| m_{1}|/ 2 } ( 1 - \eta_{1}^{ 2 } )^{|m_{3}|/2} = (1 - \eta_{1}^{2})^{|\sigma|/2} (1 - \eta_{1}^{2})^{(|m_{1} | + |m_{3}| - |\sigma|)/ 2}  \\  = ( 1 - \eta_{ 1}^{ 2 } )^{ |\sigma| / 2 } \sum_{ c_{1} = 0 }^{ (|m_{1}| + |m_{3}| - |\sigma|)/2}  ( - )^{ c_{1} } \binom{ \frac{|m_{1}| + | m_{3}| - |\sigma|}{ 2 } }{ c_{1}} \eta_{ 1 }^{2 c_{ 1 } } ; \\ 
   \\   \eta^{g} e^{- \beta \eta } = \left(  \frac{ - \partial}{ \partial \beta }    \right)^{g} e^{- \beta \eta } ; \end{multline*} 

\begin{multline*} \int_{ - 1 }^{ 1 } d \eta_{ 1 } P_{ \mu}^{ |\sigma|} ( \eta_{ 1 } ) ( 1 - \eta_{ 1 }^{2})^{ (|m_{1}| + |m_{ 3 }|)/ 2 } \eta_{ 1 }^{ k } e^{ -\beta_{ 1 } \eta_{ 1 } } =   \sum_{ c_{ 1 } = 0 }^{(|m_{1}| + |m_{3}|-|\sigma|)/2} (-)^{c_{1}} \\ \times  \binom{ \frac{|m_{1}| + | m_{3}| - |\sigma|}{ 2 } }{ c_{1}}     \left(\frac{ - \partial}{ \partial \beta_{1} } \right)^{ k + 2 c_{ 1 } }   \int_{ - 1 }^{ 1 } d \eta_{ 1 } P_{ \mu}^{ |\sigma|} ( \eta_{ 1 } ) ( 1 - \eta_{ 1 }^{2})^{| \sigma| / 2 }  e^{ -\beta_{ 1 } \eta_{ 1 } }                    \end{multline*}

                                                                                \begin{multline}  I = \left(  R^{ \sum n_{ i } + 1 } \prod_{i=1}^{4} \delta_{ i }^{ n_{ i } + 1 /2 } \right)   ( - )^{  \sigma + l_{ 3 } - | m_{ 3 } | + l_{ 4 } - | m_{ 4 } | }  \left( \frac{ 1 }{ 2 } \right)  \sum_{ a_{ 1 } = 0 }^{ n_{ 1 } - l_{ 1 } }  \sum_{ b_{ 1 } = 0 }^{ n_{ 3 } - l_{ 3 } } \\ \times   \sum_{ a_{ 2 } = 0 }^{ n_{ 2 } - l_{ 2 } } \sum_{ b_{ 2 } = 0 }^{ n_{ 4 } - l_{ 4 } }  ( -  )^{ b_{ 1 } + b_{ 2 } }  \binom{ n_{ 1 } - l_{ 1 } }{ a_{ 1 } }  \binom{ n_{ 3 } - l_{ 3 } }{ b_{ 1 } } \binom{ n_{ 2 } - l_{ 2 } }{ a_{ 2 } } \binom{ n_{ 4 } - l_{ 4 } }{ b_{ 2 } }  \\ \times \prod_{ i = 1 }^{ 4 } \left\{ \left[ \frac{ ( 2 l_{ i } + 1 ) ( l_{ i } - | m_{ i } | ) ! ( l_{ i } + | m_{ i } | ) ! }{ ( 2 n_{ i } ) ! } \right]^{ 1 / 2  } \,  \sum_{ s_{ i } = 0 }^{ [ ( l_{ i } - | m_{ i } | ) / 2 ] } \sum_{ p_{ i } = 0 }^{ 2 s_{ i } }    \right.     \\ \times       \sum_{ q_{ i } = 0 }^{ l_{ i } - | m_{ i } | - 2 s_{ i } }     \frac{ ( - )^{ s_{ i } - q_{ 3 } - q_{ 4 } + p_{ 3 } + p_{ 4 } +(|m_{i}| + m_{i})/ 2 }  }{ 2^{ l_{ i } } l_{ i } ! } \binom{2 l_{ i } - 2 s_{ i } }{ l_{ i } - | m_{ i } | - 2 s_{ i } } \binom{l_{ i } }{ s_{ i } } \\ \times   \binom{ l_{ i } - | m_{ i } | - 2 s_{ i } }{ q_{ i } }  \binom{ 2 s_{ i } }{ p_{ i } }          \}\  \sum_{ c_{ 1 } = 0 }^{ ( | m_{ 1 } | + | m_{ 3 } | - | \sigma | ) / 2 }  \sum_{ c_{ 2 } = 0 }^{ ( | m_{ 2 } | + | m_{ 4 } | - | \sigma | ) / 2 }        \\  \times       \sum_{ d_{ 1 } = 0 }^{ ( | m_{ 1 } | + | m_{ 3 } | - | \sigma | ) / 2 }         \sum_{ d_{ 2 } = 0 }^{ ( | m_{ 2 } | + | m_{ 4 } | - | \sigma | ) / 2 }   ( - )^{ c_{ 1 } + c_{ 2 } + d_{ 1 } + d_{ 2 } }   \\  \times             \binom{ ( | m_{ 1 } | + | m_{ 3 } | - | \sigma | ) / 2  }{ c_{ 1 } }  \binom{ ( | m_{ 1 } | + | m_{ 3 } | -| \sigma | ) / 2 }{ d_{ 1 } }              \binom{ ( | m_{ 2 } | + | m_{ 4 } | - | \sigma | ) / 2 }{ c_{ 2 } }  \\ \times         \binom{ ( | m_{ 2 } | + | m_{ 4 } | - | \sigma | ) / 2 }{ d_{ 2 } }     \sum_{\mu = |\sigma|}^{\infty}   (2 \mu + 1) \left[ \frac{ (\mu - | \sigma | )|} {( \mu + |\sigma|)! }    \right]^{2}      \\ \times  \left[  ( \frac{ - \partial}{ \partial \beta_{1}} )^{g_{1} }\int_{ - 1}^{ 1}  P_{\mu}^{|\sigma|} (\eta_{1}) e^{-\beta_{1} \eta_{1}} ( 1 - \eta_{1}^{2})^{|\sigma|/2} d \eta_{1} \right]     \\ \times  \left[  ( \frac{ - \partial}{ \partial \beta_{2}} )^{g_{2} }\int_{ - 1}^{ 1}  P_{\mu}^{|\sigma|} (\eta_{2}) e^{-\beta_{2} \eta_{2}} ( 1 - \eta_{2}^{2})^{|\sigma|/2} d \eta_{2} \right]    \\                         \times    \left( \frac{ - \partial}{ \partial \alpha_{ 1 } } \right)^{ r_{ 1 } }     \left( \frac{ - \partial}{ \partial \alpha_{ 2 } } \right)^{ r_{ 2 } }  \int_{ 1 }^{ \infty } \int_{ 1 }^{ \infty } d \xi_{ 1 } d \xi_{ 2 }  \\ \times     ( \xi_{ 1 }^{ 2 } - 1 )^{ | \sigma | / 2  }   ( \xi_{ 2 }^{ 2 } - 1 )^{ | \sigma | / 2  } P_{ \mu}^{ | \sigma | } ( \xi_{ 1 < 2 } )    Q_{ \mu }^{ | \sigma | } ( \xi_{ 2 > 1 } ) e^{ - \alpha_{ 1 } \xi_{ 1 } } e^{ - \alpha_{ 2 } \xi_{ 2 } }                 \end{multline}

    The square brackets in the upper limit of summation over $s_{ i } $ is $ \frac{ 1 }{ 2 }  ( l_{ i } - | m_{ i } |  ) $  or   $ \frac{ 1 }{ 2 }  ( l_{ i } - | m_{ i }  |  - 1  ) $    whichever is an integral.     

\begin{multline*}  g_{ 1 } = p_{ 1 } + p_{ 3 } + q_{ 1 } + q_{ 3 } + a_{ 1 } + b_{ 1 } + 2 c_{ 1 } , \\   g_{ 2 } = p_{ 2 } + p_{ 4 } + q_{ 2 } + q_{ 4 } + a_{ 2 } + b_{ 2 } + 2 c_{ 2 } , \\                                                     r_{ 1 } = 2 ( s_{ 1 } + s_{ 3 } + d_{ 1 } ) + q_{ 1 } + q_{ 3 } - p_{ 1 } - p_{ 3 } + n_{ 1 } + n_{ 3 } - l_{ 1 } - l_{ 3 } - a_{ 1 } - b_{ 1 }, \\             r_{ 2 } = 2 ( s_{ 2 } + s_{ 4 } + d_{ 2 } ) + q_{ 2 } + q_{ 4 } - p_{ 2 } - p_{ 4 } + n_{ 2 } + n_{ 4 } - l_{ 2 } - l_{ 4 } - a_{ 2 } - b_{ 2 }  ,            \end{multline*}

\begin{multline*}   \alpha_{ 1 } = \frac{ R }{ 2 } ( \delta_{ 1 } + \delta_{ 3 } )  ; \quad  \alpha_{ 2 } = \frac{ R }{ 2 } ( \delta_{ 2 } + \delta_{ 4 } ) ;   \quad \beta_{ 1 } = \frac{ R }{ 2 } ( \delta_{ 1 } - \delta_{ 3 } ) ; \\  \quad      \beta_{ 2 } = \frac{ R }{ 2 } ( \delta_{ 2 } - \delta_{ 4 } ) ; \quad   W = R^{ \sum n_{ i } + 1 } \Pi \delta_{ i }^{ n_{ i } + 1 /2 } \\ W = \delta_{ 1 } ( \alpha_{ 1 } + \beta_{ 1 } )^{ n_{ 1 } - 1 / 2 } ( \alpha_{ 1 } - \beta_{ 1 } )^{ n_{ 3 } + 1/2 } ( \alpha_{ 2 } + \beta_{ 2 } )^{ n_{ 2 } + 1 / 2 } (\alpha_{ 2 } - \beta_{ 2 } )^{ n_{ 4 } + 1 / 2 } \\   W = \frac{ 1 }{ R } ( \alpha_{ 1 } + \beta_{ 1 } )^{ n_{ 1 } + 1 / 2 } ( \alpha_{ 1 } - \beta_{ 1 } )^{ n_{ 3 } + 1/2 } ( \alpha_{ 2 } + \beta_{ 2 } )^{ n_{ 2 } + 1 / 2 } (\alpha_{ 2 } - \beta_{ 2 } )^{ n_{ 4 } + 1 / 2 }       \quad \quad \quad   \\     \text{For the Hybrid integral } =  \int \int d \tau_{1} d \tau_{2}  \Phi_{ 1 a } ( 1 ) \Phi_{ 3 a } ( 1 ) \frac{ 1 }{ r_{ 1 2 } } \Phi_{ 2 a } ( 2 ) \Phi_{ 4 b } ( 2 ); \quad \\ \quad     \alpha_{ 1 } = \beta_{1} =  \frac{ R }{ 2 } ( \delta_{ 1 } + \delta_{ 3 } )  ; \quad  \alpha_{ 2 } = \frac{ R }{ 2 } ( \delta_{ 2 } + \delta_{ 4 } ) ;   \quad      \beta_{ 2 } = \frac{ R }{ 2 } ( \delta_{ 2 } - \delta_{ 4 } ) ; \quad \\ \quad                                         \text{For the Coulomb integral } =  \int \int d \tau_{1} d \tau_{2}  \Phi_{ 1 a } ( 1 ) \Phi_{ 3 a } ( 1 ) \frac{ 1 }{ r_{ 1 2 } } \Phi_{ 2 b } ( 2 ) \Phi_{ 4 b } ( 2 ) ;  \quad  \\ \quad  \alpha_{ 1 } = \beta_{ 1 } =\frac{ R }{ 2 } ( \delta_{ 1 } + \delta_{ 3 } )  ; \quad  \alpha_{ 2 } = - \beta_{ 2 } = \frac{ R }{ 2 } ( \delta_{ 2 } + \delta_{ 4 } ) ;    \quad \end{multline*}
\,  \,  \,  \,  \,  \,  \,  \,  \,  \,  \,  \,  \,  \,  \,  \,  \,\,  \,  \,  \,  \,  \,  \,  \,  \,  \,  \,  \,  \,  \,  \,  \,  \,  \,  \,  \,  \,  \,  \,  \,  \,  \,  \,  \,  \,  \,  \,  \,  \,  \,  \,  \,  \,  \,  \,  \,  \,  \,  \,  \,  \,  \,  \,  \,  \,  \,  \,  \,  \,  \,  \,  \,  \,  \,  \,  \,  \,  \,  \,  \,  \,  \,  \,  \,  \,  \,  \,  \,  \,  \,  \,  \,  \,  \,  \,  \,  \,  \,  
   
                              These changes result in changes in the set of  $ g_{ 1 k } , g_{ 2 k }, r_{ 1 k }. r_{ 2 k }. C_{ k } $  and the coefficients at the beginning of equation 1.

\begin{multline}   ( \frac{ - \partial}{ \partial \beta_{1}} )^{g }\int_{ - 1}^{ 1}  P_{\mu}^{|\sigma|} (\eta_{1}) e^{-\beta_{1} \eta_{1}} ( 1 - \eta_{1}^{2})^{|\sigma|/2} d \eta_{1}     \\  =   \left(  \frac { - \partial }{ \partial \beta } \right)^{ g } ( - )^{ \mu }  \sqrt{ 2 \pi } \  \frac{ (\mu + | \sigma|)!}{ \mu - | \sigma|)!}     \frac{ I_{ \mu + 1 / 2 } ( \beta ) }{ \beta^{ | \sigma | + 1 / 2 } }  \\ \text{Use has been made of formulas from Ref. 15 (7.321 pg. 830}\\ \text{, 8.936.2 pg. 1031, 8.406.1 pg. 952 )  and Ref. 14 ( 6.1.18 pg. 256 ).}         \end{multline}                                                                           \begin{multline*} \text{   Using the finite representation of     the modified spherical Bessel} \\ \text{ function of the first kind $I_{\mu +1/2} (\beta)$ ( Ref. 14 , 10.2.9 )    and differentiating } \\                                \text{ If } \beta \neq  0 \text{  then Eq. 2 is :  }  = \frac{ ( - )^{ 2 \mu + 1 } }{ \beta^{ | \sigma | + 1 } } \frac{ ( \mu + |\sigma| ) ! }{ (\mu - |\sigma|)!}  \sum_{ k = 0 }^{ \mu } \frac{ \mu ! }{ ( \mu - k ) !}  \binom{ \mu + k }{ k }  \\ \times   \frac{ 1 }{ ( 2 \beta )^{ k } }   \sum_{ j = 0 }^{ g } \binom{ g }{ j } \binom{ k + | \sigma | + j }{ j } \frac{ j ! }{ \beta^{ j } }  \left[ ( - )^{ k + j + g + \mu + 1 } e^{ \beta }  + e^{ - \beta } \right]   \\  \text{ In practice this latter formula was not used.  The ascending } \\ \text{series  ( Ref. 14, 10.2.5 ) representation and differentiating} \\ \text{ proved to be more accurate ( less roundoff errors ) } \\  \text{ for } \beta \neq 0 \text{ and } 2 k + \mu - | \sigma | \geq g ,  \text{ then Eq. 2 is:   } \\  =  ( - )^{\mu +  g } 2^{ \mu + 1 } \frac{ ( \mu + |\sigma|)!}{( \mu - |\sigma|)!} \beta^{ \mu - | \sigma | - g } \\ \times  \sum_{ k = 0 }^{ \infty } \frac{ \beta^{ 2 k } \binom{ \mu + 2 k - | \sigma | }{ g } \frac{ g ! }{ k ! \mu ! ( k + 1 ) ! } }{ \binom{ 2 \mu + 2 k + 1 }{ \mu + k } \binom{ \mu + k + 1 }{ \mu } } \\         \text{ If } \beta = 0 \text{  then Eq. 2 is : } \\  =  \frac{ ( - )^{2 \mu - | \sigma | } 2^{ \mu + 2 } \frac{ ( \mu + |\sigma|)!}{\mu - |\sigma|)!}  \binom{ 1 + g + | \sigma | } {\frac{ g + | \sigma | - \mu }{ 2 } }  }{ ( 1 + | \sigma | ) ! \binom{ g + | \sigma | + 1 }{ g } \binom{ 2 + g + | \sigma | + \mu }{ 1 + \frac{ g + | \sigma | + \mu }{ 2 } }  }                            \end{multline*}   
This is arrived at either through the limit $\beta \rightarrow $ 0 of the ascending series form of the original integral or by use of Ref. 15 ( 7.132.5 pg. 799 ) and Ref. 14 ( 6.1.18) .   

\begin{multline} \text{ Reduction of the remaining integral}\\ \text{ from Equation 1 into it's component parts:}\\                 
   \int_{ 1 }^{ \infty } \int_{ 1 }^{ \infty }  d \xi_{ 1 } d \xi_{ 2 }  \\ \times     ( \xi_{ 1 }^{ 2 } - 1 )^{ | \sigma | / 2  }   ( \xi_{ 2 }^{ 2 } - 1 )^{ | \sigma | / 2  } P_{ \mu}^{ | \sigma | } ( \xi_{ 1 < 2 } )    Q_{ \mu }^{ | \sigma | } ( \xi_{ 2 > 1 } ) e^{ - \alpha_{ 1 } \xi_{ 1 } } e^{ - \alpha_{ 2 } \xi_{ 2 } } \\ =    \int_{ 1 }^{ \infty }      Q_{ \mu }^{ | \sigma | } ( \xi_{ 2  } )    ( \xi_{ 2 }^{ 2 } - 1 )^{ | \sigma | / 2  }  e^{ - \alpha_{ 2 } \xi_{ 2 } }    d \xi_{ 2 }   \int_{ 1 }^{ \xi_{2}  }    P_{ \mu}^{ | \sigma | } ( \xi_{ 1  } )     ( \xi_{ 1 }^{ 2 } - 1 )^{ | \sigma | / 2  }   e^{ - \alpha_{ 1 } \xi_{ 1 } }    d \xi_{ 1 } \\   +    \int_{ 1 }^{ \infty }      P_{ \mu }^{ | \sigma | } ( \xi_{ 2  } )    ( \xi_{ 2 }^{ 2 } - 1 )^{ | \sigma | / 2  }  e^{ - \alpha_{ 2 } \xi_{ 2 } }    d \xi_{ 2 }   \int_{ \xi_{2} }^{ \infty}     Q_{ \mu}^{ | \sigma | } ( \xi_{ 1  } )     ( \xi_{ 1 }^{ 2 } - 1 )^{ | \sigma | / 2  }   e^{ - \alpha_{ 1 } \xi_{ 1 } }    d \xi_{ 1 } \\ = \left[ 1 + O_{per} \binom{1}{2} \right]      \int_{ 1 }^{ \infty }      Q_{ \mu }^{ | \sigma | } ( \xi_{ 2  } )    ( \xi_{ 2 }^{ 2 } - 1 )^{ | \sigma | / 2  }  e^{ - \alpha_{ 2 } \xi_{ 2 } }    d \xi_{ 2 } \\ \times   \int_{ 1 }^{ \xi_{2}  }    P_{ \mu}^{ | \sigma | } ( \xi_{ 1  } )     ( \xi_{ 1 }^{ 2 } - 1 )^{ | \sigma | / 2  }   e^{ - \alpha_{ 1 } \xi_{ 1 } }    d \xi_{ 1 } \\  \text{  The order of integration was changed using Ref. 15 ( 4.611 pg. 615 ).} \end{multline}    
\begin{multline*}  Q_{\mu}^{|\sigma|}(z) ( z^{ 2 } -1 )^{|\sigma|/2} = \frac{1}{2}  P_{\mu}^{|\sigma|}( z ) (z^{2} -1 )^{|\sigma|/2} ln ( \frac{z + 1}{z - 1 } ) \\  + \frac{1}{2} \sum_{\kappa =1}^{|\sigma|}\binom{|\sigma|}{\kappa} P_{\mu}^{|\sigma|-\kappa} (z) (z^{2} - 1)^{(|\sigma| - \kappa )/2} (- )^{ \kappa - 1 } (\kappa - 1 )!  [ (z-1)^{\kappa} - ( z + 1)^{\kappa} ]  \\  - \sum_{ j = 0 }^{[(\mu - 1 - |\sigma|)/2]}  \frac{2 \mu - 4 j - 1 }{( 2 j + 1)( \mu - j )} P_{ \mu - 2 j - 1 }^{|\sigma|} ( z ) ( z^{ 2 } - 1 )^{|\sigma|/2}  \\ \text{ Derived from definitions ( Ref. 14. , 8.6.19, 8.6.7 ) }                           \end{multline*}
   
\begin{multline*}  \text{Eqn. 3  } = [ 1 + O_{per} \binom{1}{2} ] [ A + B + C ] ;\\
		A = \frac{1}{4} \int_{1}^{\infty} dz [ \frac{1}{z-1} - \frac{1}{z+1} ] \int_{1}^{z}P_{\mu}^{|\sigma|}(x)(x^{2} -1)^{|\sigma|/2}e^{-\alpha_{2} x} dx \\ \times    \int_{1}^{z} P_{\mu}^{|\sigma|}(y) (y^{2} - 1)^{|\sigma|/2} e^{-\alpha_{1} y } dy   \\    \text{ Integration by parts was used in A.} \\ \text{Integrating from 1 to z :}  \\ A =     \left[ \frac{ ( \mu + | \sigma | ) ! }{  2^{\mu + 1} \mu ! } \right]^{ 2 }  \sum_{ k = 0}^{ \left[ \frac{ \mu + | \sigma | }{ 2 } \right] }  ( - )^{ k } \binom{ 2 \mu - 2 k }{ \mu - | \sigma | } \binom{ \mu }{ k }    (\frac{-\partial}{\partial \alpha_{2}})^{\mu + |\sigma| - 2 k } \\ \times    \sum_{ p = 0}^{ \left[ \frac{ \mu + | \sigma | }{ 2 } \right] }  ( - )^{ p } \binom{ 2 \mu - 2 p }{ \mu - | \sigma | } \binom{ \mu }{ p }  (\frac{-\partial}{\partial \alpha_{1}})^{\mu + |\sigma| - 2 p }  \\ \times \int_{1}^{\infty} dz [ \frac{1}{z - 1} - \frac{1}{z + 1} ] [ e^{-\alpha_{1}} - e^{-\alpha_{1}z} ]  [ e^{-\alpha_{2}} - e^{-\alpha_{2}z} ]  \\ \text{ Using Ref. 16 ( 860.22 pg. 230) and Ref. 14 ( 5.1.39, 5.1.1 ) } \\  \int_{1}^{\infty} dz [ \frac{1}{z - 1} - \frac{1}{z + 1} ] [ e^{-\alpha_{1}} - e^{-\alpha_{1}z} ]  [ e^{-\alpha_{2}} - e^{-\alpha_{2}z} ] \\ =  \int_{0}^{\infty} [e^{-\alpha_1} -  e^{-\alpha_1(1 + 2z)} ] [ e^{-\alpha_2} - e^{-\alpha_2 ( 1 + 2z)} ] \frac{dz}{z} \\ -    \int_{1}^{\infty} [e^{-\alpha_1} -  e^{-\alpha_1( 2z - 1)} ] [ e^{-\alpha_2} - e^{-\alpha_2 (  2z - 1)} ] \frac{dz}{z} \\ = e^{-(\alpha_1 + \alpha_2)} [ \ln (\frac{2 \alpha_{1} \alpha_{2} }{\alpha_{1} + \alpha_{2}} + \gamma + e^{2 \alpha_{1} }E_{1}(2 \alpha_{1})  \\  + e^{2 \alpha_{2}}E_{1}(2 \alpha_{2}) - e^{2(\alpha_{1} + \alpha_{2})}E_{1}(2\alpha_{1} + 2\alpha_{2})]     \end{multline*}                 \begin{multline*}                           B = \sum_{\kappa = 1}^{|\sigma|}  \binom{|\sigma|}{\kappa} (\kappa - 1)!(-)^{\kappa} \sum_{j=0}^{[(\kappa-1)/2]} \binom{\kappa}{\kappa - 2 j - 1} \\ \times                         \int_{1}^{\infty} P_{\mu}^{|\sigma| - \kappa} (z) (z^{2} - 1)^{(|\sigma|-\kappa)/2} z^{\kappa - 2 j - 1} e^{-\alpha_{2} z} dz \\ \times \int_{1}^{z} P_{\mu}^{|\sigma|} (x) (x^{2} - 1)^{|\sigma|/2} e^{-\alpha_{1} x } dx  ;\\ \text{ in B, use was made of : } \\       (z+1)^{\kappa} - (z-1)^{\kappa} = 2 \sum_{j=0}^{[\frac{\kappa - 1}{2}]} \binom{\kappa}{\kappa - 2 j - 1} z^{\kappa - 2 j - 1} ; \\  C =  - \sum_{j=0}^{[(\mu - 1 - |\sigma|)/2]}  \frac{2 \mu - 4 j - 1}{(2 j + 1)(\mu - j)}                                           \int_{1}^{\infty} P_{\mu - 2 j - 1}^{|\sigma|} (z) (z^{2} - 1)^{|\sigma|/2}  e^{-\alpha_{2} z} dz \\ \times \int_{1}^{z} P_{\mu}^{|\sigma|} (x) (x^{2} - 1)^{|\sigma|/2} e^{-\alpha_{1} x } dx ; \\  \text{ A representation of $P_{\mu}^{|\sigma|} (z) (z^{2} - 1 )^{|\sigma|/2} $ } \\ \text{  is used by combining 8.6.6,8.6.18,8.2.5 (Ref. 14),}\\  P_{\mu}^{|\sigma|} (z)   = \frac{( \mu + |\sigma|)! (z^2 - 1 )^{-|\sigma|/2 } } {2^{\mu } \mu ! ( \mu - |\sigma|) ! } \frac{d ^{\mu - |\sigma|}}{dz^{\mu - |\sigma|} } (z^{2} -1 )^{\mu}   \\  \text{ using  binomial expansion and differentiating}  ; \\  P_{\mu}^{|\sigma|} (z) (z^{2} - 1 )^{|\sigma|/2}  = \frac{( \mu + |\sigma|)!}{2^{\mu } \mu !}  \sum_{p=0}^{[\frac{\mu + |\sigma|}{2}]} ( - )^{p} \binom{2\mu - 2p}{\mu - |\sigma|} \binom{ \mu}{p} z^{\mu + |\sigma| - 2 p }     \end{multline*}  
 \begin{multline*}    \left( \frac{ - \partial}{ \partial \alpha_{ 1 } } \right)^{ r_{ 1 } }     \left( \frac{ - \partial}{ \partial \alpha_{ 2 } } \right)^{ r_{ 2 } }  \int_{ 1 }^{ \infty } \int_{ 1 }^{ \infty } d \xi_{ 1 } d \xi_{ 2 }   \\      \times  ( \xi_{ 1 }^{ 2 } - 1 )^{ | \sigma | / 2  }   ( \xi_{ 2 }^{ 2 } - 1 )^{ | \sigma | / 2  } P_{ \mu}^{ | \sigma | } ( \xi_{ 1 < 2 } ) Q_{ \mu }^{ | \sigma | } ( \xi_{ 2 > 1 } ) e^{ - \alpha_{ 1 } \xi_{ 1 } } e^{ - \alpha_{ 2 } \xi_{ 2 } }   \\   =  \frac{ e^{ - ( \alpha_{ 1 } + \alpha_{ 2 } ) } }{ 2^{ 2 \mu + 1 } }  \left[ \frac{ ( \mu + | \sigma | ) ! }{ \mu ! } \right]^{ 2 }  \sum_{ k = 0}^{ \left[ \frac{ \mu + | \sigma | }{ 2 } \right] }  ( - )^{ k } \binom{ 2 \mu - 2 k }{ \mu - | \sigma | } \binom{ \mu }{ k }  \\ \times    \sum_{ p = 0}^{ \left[ \frac{ \mu + | \sigma | }{ 2 } \right] }  ( - )^{ p } \binom{ 2 \mu - 2 p }{ \mu - | \sigma | } \binom{ \mu }{ p } \left\{   \right. \sum_{ n_{ 1 } = 0 }^{ k_{ 1 } } \frac{ k_{ 1 } ! }{ ( k_{ 1 } - n_{ 1 } ) !  \,  \alpha_{ 1 }^{ n_{ 1 } + 1 } }   \sum_{ n_{ 2 } = 0 }^{ k_{ 2 } } \frac{ k_{ 2 } ! }{ ( k_{ 2 } - n_{ 2 } ) !  \,  \alpha_{ 2 }^{ n_{ 2 } + 1 } } \\ \left[      \right.  \ln \left( \frac { 2 \alpha_{ 1 } \alpha_{ 2 } }{ \alpha_{ 1 } + \alpha_{ 2 } } \right)  + ( - )^{ n_{ 1 } + n_{ 2 } + k_{ 1 } + k_{ 2 } + 1 } e^{ 2 ( \alpha_{ 1 } + \alpha_{ 2 } ) }   E_{ 1 } ( 2 \alpha_{ 1 } + 2 \alpha_{ 2 } )  \\ + ( - )^{ k_{ 1 } + n_{ 1 } } e^{ 2 \alpha_{ 1 } }  E_{ 1 } ( 2 \alpha_{ 1 } )   +  ( - )^{ k_{ 2 } + n_{ 2 } } e^{ 2 \alpha_{ 2 } }  E_{ 1 } ( 2 \alpha_{ 2 } )  +  \gamma \left. \right] \\   + \sum_{ n_{ 2 } = 1 }^{ k_{ 2 } } \frac{ 1 }{ n_{ 2 } }  \sum_{ j_{ 2 } = 0 }^{ k_{ 2 } - n_{ 2 } } \frac{ k_{ 2 } !  }{ ( k_{ 2 } - n_{ 2 } - j_{ 2 } ) !  }   \sum_{ n_{ 1 } = 0 }^{ k_{ 1 } } \left[  \right.  \frac{ - k_{ 1 } ! }{ ( k_{ 1 } - n_{ 1 } ) ! }  \frac{   1  }{ \alpha_{1 }^{ n_{ 1 } + 1 }   \alpha_{ 2 }^{ n_{ 2 } + j_{ 2 } + 1 } }   \\ + \frac{  k_{ 1 } ! }{ ( k_{ 1 } - n_{ 1 } ) ! }   ( - )^{ k_{ 2 } + n_{ 2 } + j_{ 2 } } \sum_{ t = 0 }^{ n_{ 2 } - 1 } \frac{  2^{ n_{ 2 } - t - 1 } }{ ( n_{ 2 } - t - 1 ) ! \,  \alpha_{ 1 }^{ n_{ 1 } + 1 } \alpha_{ 2 }^{ j_{ 2 } + t + 2 } } \\ + \sum_{ j_{ 1 } = 0 }^{ n_{ 1 } }  \frac{ k_{ 1 } ! }{ ( n_{ 1 } - j_{ 1 } ) ! } \left(  \right.  \binom{ k_{ 1 } + n_{ 2 } - n_{ 1 } - 1 }{ k_{ 1 } - n_{ 1 } } \frac{  1 }{ \alpha_{ 2 }^{ j_{ 2 } + 1 } ( \alpha_{ 1 } + \alpha_{ 2 } )^{ k_{ 1 } + n_{ 2 } - n_{ 1 } } \alpha_{ 1 }^{ j_{ 1 } + 1 } }  \\ + \sum_{ t = 0 }^{ n_{ 2 } - 1 } \frac{ ( - )^{ k_{ 2 } + n_{ 2 } + j_{ 2 } + 1 } 2^{ n_{ 2 } - t - 1 } }{ ( n_{ 2 } - t - 1 ) ! }  \binom{ t + k_{ 1 } - n_{ 1 } }{ t }  \frac{  1 }{ \alpha_{ 2 }^{ j_{ 2 } + 1 } ( \alpha_{ 1 } + \alpha_{ 2 } )^{ k_{ 1 } + t + 1  - n_{ 1 } } \alpha_{ 1 }^{ j_{ 1 } + 1 } }   \left.  \right) \left.  \right]    \\   + \sum_{ n_{ 2 } = 0 }^{ k_{ 2 }  }  \frac{ k_{ 2 } ! }{ ( k_{ 2 } - n_{ 2 } ) ! } \sum_{ n_{ 1 } = 1 }^{ k_{ 1 } }  \frac{ 1 }{ n_{ 1 } } \sum_{ j_{ 1 } = 0 }^{  k_{ 1 } - n_{ 1 } } \frac{ k_{ 1 } ! }{ ( k_{ 1 } - n_{ 1 } - j_{ 1 } ) ! }  \left[   \right.   \frac{ 1 }{ \alpha_{ 1 }^{ j_{ 1 } + 1 } \alpha_{ 2 }^{ n_{ 2 } + 1 } ( \alpha_{ 1 } + \alpha_{ 2 } )^{ n_{ 1 } } }  \\ -  \frac{ 1 }{ \alpha_{ 1 }^{ n_{ 1 } + j_{ 1 } + 1 }  \alpha_{ 2 }^{ n_{ 2 } + 1 } }  + \sum_{ t = 0 }^{ n_{ 1 } - 1 } \frac{ ( - )^{ k_{ 1 } + j_{ 1 } + n_{ 1 } } 2^{ n_{ 1 } - 1 - t } }{ ( n_{ 1 } - t - 1 ) !  }   \left( \right.  \frac{ 1 }{ \alpha_{ 1 }^{ j_{ 1 } + t + 2 }  \alpha_{ 2 }^{ n_{ 2 } + 1 } }   \\ +  \frac{  ( - )^{ k_{ 2 } + n_{ 2 } + 1 } }{ ( \alpha_{ 1 } + \alpha_{ 2 } )^{ t + 1 } \alpha_{ 1 }^{ j_{ 1 } + 1 }  \alpha_{ 2 }^{ n_{ 2 } + 1 } }  \left. \right) \left.  \right] \left.  \right\}   +  \frac{ e^{ - ( \alpha_{ 1} + \alpha_{ 2 } ) }  ( \mu + | \sigma | ) !  }{ 2^{ 2  \mu }  \mu  !  }  \sum_{ k = 0 }^{ [ \frac{ \mu + | \sigma | }{ 2 } ] } ( - )^k \binom{ 2 \mu - 2 k }{ \mu - | \sigma | }   \\  \times   \binom{ \mu }{ k } \left[  \right.    \sum_{ \kappa  = 1 }^{ | \sigma | }  \frac{ ( - )^{ \kappa  }  | \sigma | !  }{ \kappa  } \binom{ \mu + | \sigma | - \kappa }{ \mu }  \sum_{ j = 0 }^{ [ \frac{ \kappa - 1 }{ 2 } ] } \binom{ \kappa }{ \kappa - 2 j - 1 }  \sum_{ n = 0 }^{ [ \frac{ \mu + | \sigma | - \kappa }{ 2 } ] } ( - )^{ n }    \\  \times    \binom{ 2 \mu - 2 n }{ \mu - | \sigma | + \kappa }  \binom{ \mu }{ n }        -  \sum_{ j = 0 }^{ [ \frac{ \mu - | \sigma | - 1 }{ 2 } ] }  \frac{ ( 2 \mu - 4 j - 1 ) ( \mu - 2 j - 1 + | \sigma | ) !  \, 2^{ 2 j + 1 } }{ ( 2 j + 1 ) ( \mu - j ) ( \mu - 2 j - 1 ) ! }   \\ \times \sum_{ n = 0 }^{ [ \frac{ \mu + | \sigma | - 2 j - 1 }{ 2 } ] }  ( - ) ^{ n } \binom{ 2 ( \mu - 2 j - 1 - n ) }{ \mu - 2 j - 1 - | \sigma | } \binom{ \mu - 2 j - 1 }{ n } \left.  \right]               \end{multline*}                                                                              \begin{multline}          \times \left[ \right.  \sum_{ n_{ 1 } = 0 }^{ k_{ 1 }^{ o } } \frac{ k_{ 1 }^{ o } ! }{ ( k_{ 1 }^{ o } - n_{ 1 } ) ! \, }  \sum_{ n_{ 2 } = 0 }^{ f_{ 2 } } \binom{ n_{ 1 } + f_{ 2 } - n_{ 2 } }{ f_{ 2 } - n_{ 2 } }   \\ \times       \sum_{ j_{ 2 } = 0 }^{ n_{ 2 } }  \frac{ f_{ 2 } !  }{ ( n_{ 2 } - j_{ 2 } ) ! \, \alpha_{ 2 }^{ j_{ 2 } + 1 } ( \alpha_{ 1 } + \alpha_{ 2 } )^{ f_{ 2 } + n_{ 1 } - n_{ 2 } + 1 } }   \\   + \sum_{ n_{ 2 } = 0 }^{ k_{ 2 } }  \frac{ k_{ 2 } ! }{ ( k_{ 2 } - n_{ 2 } ) ! \, }  \sum_{ n_{ 1 } = 0 }^{ f_{ 1 } } \binom{ n_{ 2 } + f_{ 1 } - n_{ 1 } }{ f_{ 1 } - n_{ 1 } }  \\ \times    \sum_{ j_{ 1 } = 0 }^{ n_{ 1 } }  \frac{ f_{ 1 } ! }{ ( n_{ 1 } - j_{ 1 } ) ! \, \alpha_{ 1 }^{ j_{ 1 } + 1 } ( \alpha_{ 1 } + \alpha_{ 2 } )^{ f_{ 1 } + n_{ 2 } - n_{ 1 } + 1 } }  \left.  \right]  .  \\  E_{ 1 } ( x )  =  \text{ Exponential integral  ( 5.1.1. Ref. 14 ) }  \\    \gamma = \text{ Euler's constant  ( table 1.1 Ref. 14 ) }   \\ k_{ 1 } = \mu + | \sigma | - 2 p + r_{ 1 }  , \quad  k_{ 2 } = \mu + | \sigma | - 2 k + r_{ 2 } , \\  k_{ 1 }^{ o } = \mu + | \sigma | - 2 k + r_{ 1 } , \quad  f_{ 1 } = \mu + | \sigma | - 2 ( n + j ) - 1 + r_{ 1 } , \\ f_{ 2 } = \mu + | \sigma | - 2 ( n + j ) - 1 + r_{ 2 }  ,                                            \end{multline}     
\begin{center} GRAPHICAL ENUMERATION \end{center}

Knowing which integrals to evaluate with an n electron correlated wavefunction is an important combinatorial problem.  We do know that the maximum number of correlation terms in the wavefunction is $n( n - 1 )/2$.  The complexity of the matrix element of an operator involving $r_{i j }$ is no more complicated than this, since all the $r_{i j }$ terms in the operator coincide with some of the $n(n - 1)/2$   $  r_{ i j } $ terms of the wavefunction.  For n electrons there are possibilities of connections by a maximum of $n(n - 1)/2 $ lines  ( interelectronic distances ) in graph theory and a minimum of $n - 1$ lines.  If connected graphs are labelled by $(n,m)$, where n is the number of electrons and m is the number of lines, there are $(n^{2} - 3 n + 4)/2 $    $ (n,m)$ labels for a given n.  Each   $(n,m)$ corresponds to one or more configurations or graphs.  Asymptotically there are $\frac{2^{n(n-1)/2} }{n !} $ connected graphs for n $\text{points}^{17}$.  If one tries to generate the $(n,m)$ configurations, , given n and m, one possibility is to label the n electrons in some way and form the composition: $ (m_{1},m_{2},....,m_{n})$ where the $m_{i}$ ( degree of the point or vertex i ) counts the number of lines incident to point i and the sum of all the $m_{i}$ is $ 2 m $.  Beginning with $(5,6)$ there are more than one graph for a given composition ; also , not all compositions are graphical.  There are in general more than one graph for a given n and m, the number of $(n,m)$ graphs is denoted by $C(n,m)$.  For n points, $\text{Riddell}^{18-21}$ , used the $\text{Polya}^{20}$ counting theorem(which involves computing all the partitions of n ) to form the counting series for the total number of connected graphs and for the total number of graphs (connected and disconnected).  He used the product theorem to relate the counting series for the connected graphs to the series for all the graphs.  $C(n,m)$ can be obtained through this relations.  The most general terms for the generating function ( a polynomial whose coeffiecients are the $C(n,m)$) is not available in explicit form.  The numbers and diagrams for connected graphs with up to six points are readily $\text{available}^{19}$.  Graphical enumeration problems occur in enumerating terms of the Mayer cluster integral $\text{expansion}^{18,23,24}$ as well as  in enumeration of the Feynman-Dyson $\text{graphs}^{25}$.  We wish to point out that this problem also exists for correlation integrals.  Except for the case of a loop ( e.g. $r_{12}r_{23}r_{31}$ ) , $\text{L-RCM }^{26}$ is a programming method that claims to find connected components in undirected graphs.   $\text{These}^{27, 28}$ formulas can be used, where appropriate, to evaluate integrals over Slater orbitals to obtain $\text{explicitly}^{29, 30}$ correlated wavefunctions.  Relativistic effects can be $\text{included}^{31}$.  
\begin{center} ACKNOWLEDGEMENTS\end{center}
The advice and encouragement of Prof. Charles L. Schwartz are belatedly acknowledged as well as the hospitality of the Physics Dept. and the assistance of Prof. N. Sherman.                                                                      \begin{center} REFERENCES\end{center}  
                                 1.  E.V.Rothstein, Phys.Rev, \underline{A3} ,1581 (1971) . \\  2.  F.E.Harris and H.H. Michels, Adv. in Chem.Phys.\underline{13},205(1967).  \\ 3.  M. Geller, J. Chem. Phys. \underline{41},4006 (1964) \\ 4.  H.J. Silverstone, J. Chem. Phys. \underline{48}, 4098 (1968). \\ 5.  H.J. Silverstone, J. Chem. Phys. \underline{45}, 4337 (1966).  \\ 6.  D.M. Silver, J. Math. Phys. \underline{12}, 1937 (1971).  \\ 7.  A.C. Wahl, P.E. Cade and C.C.J. Roothaan, J. Chem. Phys. \underline{41}, 2578 (1964). \\ 8.  K.O. Ohata and K. Reudenberg, J. Math. Phys. \underline{7}, 547 (1966). \\ 9.  F.E. Neumann, J. Reine Angew. Math (Crelle) \underline{37}. 21 (1848).  \\ 10.  J. Podolanski, Annalen der Physik \underline{10}, 868 (1931). \\ 11.  H.J. Silverstone and K.G.Kay, J. Chem. Phys. \underline{48}, 4108 (1968).  \\ 12.  W. Kolos and L. Wolniewicz , J. Chem. Phys. \underline{41}. 3663 (1964). \\ 13.  T. Zivkovic and J.N. Murrell, Theor. Chim. Acta \underline{21}, 301 (1971).  \\     
                14.  M. Abramowitz and I.A.Stegun, editors, \textit{Handbook of Mathematical Functions}, ( Dover, New York) \\   15.  I.S. Gradshteyn and I.M. Ryzhik, \textit{ Table of Integrals, Series and Products } , (Academic Press, N.Y., 1965).  \\  16.  H.B. Dwight, \textit{Tables of Integrals and Other Mathematical Data}, ( The Macmillan Co., N.Y.,1961).  \\ 17.  G.W. Ford and G.E. Uhlenbeck, Proc. Nat. Acad. Sci. (U.S.) , \underline{43}, 163 (1957).  \\  18.  R.J. Riddell and G.E. Uhlenbeck, J. Chem. Phys. \underline{21}, 2056 (1953).  \\         19.  G.E. Uhlenbeck and G.W.Ford, "The Theory of Linear Graphs with Applications to the Theory of the Virial Development of the Properties of Gases" in \textit{Studies in Statistical Mechanics}, edited by J. DeBoer and G.E. Uhlenbeck (Interscience, N.Y.,1962), Vol. 2, Part B, pg. 119.  \\  20.  R.J. Riddell, "Contributions to the Theory of Condensation", dissertation, Univ. of Michigan, 1951.  \\  21.  F. Harary, Trans. Am. Math. Soc. \underline{78}, 445 (1955).  \\  22.  G. Polya, Acta. Math. \underline{68}, 145 (1937).  \\  23.  D.D. Fitts and W.R. Smith, Mol. Phys. \underline{22}, 625 (1971).  \\   24.  A. Felinski, Acta Phys. Polonica \underline{A40}, 315 (1971).  \\ 25.  C.A. Hurst, Proc. Roy. Soc. (Lon.) \underline{214}, 44 (1952).  \\ 26.  Francisco Pedroche, Miguel Rebollo, Carlos Carrascosa, Alberto Palomares,  L-Rcm: A Method to Detect Connected Components in Undirected Graphs by using the Laplacian Matrix and the RCM Algorithm (2012)  arXiv:1206.5726   \\ 27.  E.V. Rothstein, Electron Interaction Integrals over Slater Orbitals for Diatomic Molecules (2011) arXiv:1104.3650  \\ 28.  E.V. Rothstein, Two-Center Integrals for $r_{ij}^{n}$ Polynomial Correlated Wave Functiions (2011) arXiv:1104.3646  \\ 29.  E.A.G. Armour, J. Franz and  J. Tennyson, \textit{Explicitly Correlated Wavefuntions} ,(ISBN 0-9545289-4-8, Daresbury, 2006).  \\  30.  C. Hattig, W. Klopper, A. Kohn and D.P. Tew, Chem. Rev. \underline{112}, 4 (2012).  \\ 31.  Zhendong Li, Sihong Shao and Wanjian Liu, J. Chem. Phys. \underline{136},144117 (2012). \\
                                      :                                           \end{document}